\documentclass[12pt,preprint]{aastex}
\usepackage{graphicx}
\usepackage{amssymb}
\usepackage{amsmath}
\usepackage{natbib}
\usepackage{rotating}
\usepackage{url}
\usepackage{epstopdf}

\def\beqa{\begin{eqnarray}}
\def\eeqa{\end{eqnarray}}

\def\beq{\begin{equation}}
\def\eeq{\end{equation}}

\def\HI{{H{\small I }}}
\def\uv{{{$uv$}}}

\def\beqa{\begin{eqnarray}}
\def\eeqa{\end{eqnarray}}

\def\beq{\begin{equation}}
\def\eeq{\end{equation}}

\def\bseq{\begin{subequations}\begin{eqnarray}}
\def\eseq{\end{eqnarray}\end{subequations}}

\def\n{{\bf n}}

\def\N{{\textsf{\textbf{N}}}}
\def\Ne{\tilde{\textsf{\textbf{N}}}}

\def\B{{\textsf{\textbf{B}}}}
\def\Bv{{\tilde{\B}}}
\def\W{{\textsf{\textbf{W}}}}
\def\m{{\bf{m}}}
\def\u{{\bf{u}}}

\def\v{{\bf{v}}}

\def\I{{I}}

\def\Sv{{\textsf{\textbf{S}}}}

\def\sky{{\theta}}

\def\F{{\textsf{\textbf{F}}}}

\def\r{{\bf{r}}}

\def\m{{\bf{m}}}

\def\r{{\bf{r}}}
\def\a{{\bf{a}}}
\def\Fill{{\mathcal F}}

\begin{document}

\title{Enabling Next Generation Dark Energy and Epoch of Reionization Radio Observatories with the MOFF Correlator}

\author{Miguel F. Morales\altaffilmark{1}}
\email{mmorales@phys.washington.edu}
\altaffiltext{1}{University of Washington, Seattle, 98195}

\begin{abstract}
Proposed  21 cm cosmology observatories for studying the epoch of reionization ($z \approx$ 6--15)  and dark energy ($z \approx$ 0--6) envision compact arrays with tens of thousands of antenna elements. Fully correlating this many elements is computationally expensive using traditional XF or FX correlators, and has led some groups to reconsider direct imaging/FFT correlators. In this paper we develop a variation of the direct imaging correlator we call the MOFF correlator. The MOFF correlator shares the computational advantages of a direct imaging correlator, while avoiding a number of its shortcomings. In particular the MOFF correlator makes no constraints on the antenna arrangement or type, provides a fully calibrated output image including widefield polarimetry and non-coplanar baseline effects, and can be orders-of-magnitude more efficient than XF or FX correlators for compact radio cosmology arrays.
\end{abstract}

\section{Introduction}
A new generation of radio observatories are being developed to measure the history and evolution of our universe using deep radio surveys of large scale structure. The MWA (Murchison Widefield Array\footnote{\url{http://www.MWAtelescope.org/}}), LOFAR (LOw Frequency ARray\footnote{\url{http://www.lofar.org/}}), and PAPER (Precision Array to Probe Epoch of Reionization\footnote{\url{http://astro.berkeley.edu/~dbacker/eor/}}) are being built to observe the Epoch of Reionization (EoR) through large scale \HI emission at redshifts of 6--11, and much larger second generation EoR observatories such as HERA II\footnote{\url{http://reionization.org/}} are being planned. Additionally, at redshifts of 0--6 \HI structure experiments can trace large scale structure over much of the observable universe \citep{Wyithe:2007p4,Mao:2008p196,Chang:2008p439,Loeb:2008p3284}, enabling a number of precision cosmography and dark energy measurements. In particular the dark energy baryon acoustic oscillation (BAO) signal has inspired the development of a number of new instrument concepts, including CHIME (Canadian Hydrogen Intensity Mapping Experiment), CRT (Cylinder Radio Telescope), Omniscope \citep[formerly FFT Telescope,][]{Mao:2008p196}, CARPE (Cosmological Acceleration and Radio Pulsar Experiment \footnote{\url{http://www.phys.washington.edu/users/mmorales/carpe/}}), and LARC (Lunar Array for Radio Cosmology \footnote{\url{http://lunar.colorado.edu/}}). For a review of 21~cm cosmology and the associated instruments and observational challenges see \cite{MoralesWyithe}.

The \HI cosmology and second generation Epoch of Reionization instruments are similar in their instrumental characteristics, featuring compact arrangements of many thousands of antennas. The many short baselines and wide field of view maximizes the sensitivity of power spectrum measurements over all scales \citep{MoralesWyithe,Morales:2005p796}, however, it also puts enormous processing demands on the correlator system. A radio correlator measures the cross-power correlation between all antenna pairs in many narrow frequency channels, and for a modern FX correlator the computation scales as the square of the number of antennas. The correlator under construction for the MWA \citep{Lonsdale2008} requires 15.5 trillion complex multiplies and accumulations per second (Tcmacs) for 512 antennas and 31 MHz of bandwidth. For many of the proposed radio cosmology instruments this quickly scales into the peta-flop regime, making the correlator a dominant cost for these arrays. This has driven some concepts such as the Omniscope and cylinder telescopes to use  direct imaging correlators, despite their significant shortcomings. 

In this paper we introduce the MOFF (Modular Optimal Frequency Fourier) correlator concept. The MOFF correlator incorporates the gridding and calibration usually associated with post-correlation image processing into the correlation process. This work expands on direct imaging correlator concepts by \cite{Daishido:2000p3323} and Rogers (ATA Memo Series, 102), with particular emphases on data calibration and modular design.  For the compact antenna layouts of proposed radio cosmology telescopes, the MOFF correlator is very efficient, makes no constraints on the antenna placement, and produces a provably optimal data product.

%Talk about desire for large N arrays for dark energy, EoR, cosmography, and SKA, and the issues this presents for a modern FX correlator.

%Introduce FFT telescope concept, and it's shortcomings.

%Talk about how MOFF correlator will fix these issues:  calibration using software holography, computational efficiency, not limited to regular arrays, modular build-out, reduced cabling costs.

\section{Correlator approaches}
\label{MOFFSec}

The two most common forms of correlators are the XF and FX designs. The XF or lag correlator cross-correlates and time averages the time-domain electric field samples from each antenna pair using a set of time delays (`lags'), and then Fourier transforms the resulting set of correlation coefficients to obtain spectral information 
\beq
v_{ab}(f_{k}) = \F(f_{k},\Delta t_{l})\Big <E_{a}(t_{i})E_{b}(t_{i}+\Delta t_{l})\Big >_{t_{i}}.
%v_{lmk} = \sum_t \sum_{\Delta t_i}^N e^{-2\pi i k \Delta t_i/N} \left[ E_l(t)\cdot E^*_m(t + \Delta t_i) \right].
\label{XF}
\eeq
In the XF equation above, the electric field $E$ for two antennas $a,b$ at each time $t_{i}$ are multiplied together and time averaged at a set of discreet lags $\Delta t_{l}$. The lags are then Fourier transformed to form the visibilities (cross-power) between the antennas as a function of frequency $(v_{ab}(f_{k}))$. The linear algebra notation used for the Fourier transform ($\F$) indicates the space of the vector to be operated on ($\Delta t_{l}$ here) as the righthand argument and the output space (frequency $f_{k}$) as the lefthand argument.

The FX or spectral-domain correlator reverses the order of these operations, and Fourier transforms the electric field samples of each antenna to create spectral time-sequences, which are then multiplied and integrated to form the visibilities
\beq
v_{ab}(f_{k}) =\Big<E_{a}(f_{k},t_{j})E_{b}^{*}(f_{k},t_{j})\Big >_{t_{j}},
%v_{lmk} = \sum_{Nt} \left[ \sum_t^N e^{-2\pi i k t/N}E_l(t) \cdot \sum_t^N e^{-2\pi i k t/N}E^*_m(t) \right].
\label{FX}
\eeq
where we have used that a finite length (resolution) Fourier transform on a long time series creates a set of electric field signals in each frequency channel
\beq
E(f_{k},t_{j}) = \F([f_{k},t_{j}],t_{i})E(t_{i}).
\label{FFTofE}
\eeq
(Each time step within a spectral channel $t_{j}$ is longer than original timesteps $t_{i}$ by the number of frequency channels in the Fourier transform, so the electric field has the same amount of data before and after the transform. These frequency transforms are often implemented as polyphase filters to increase the spectral dynamic range.) As the time averaging in lag channels of Equation \ref{XF} is a temporal convolution, the FX operation in Equation \ref{FX} is just the application of the convolution-to-multiplication relationship of the Fourier transform.

Both XF and FX correlators are in wide production, with the EVLA using a modified XF correlator \citep{WIDAR} and the VLBA, MWA, and PAPER using FX correlators. As the number of antennas increases the FX correlator begins to have a significant computational advantage as only one cross product must be made per antenna pair \citep[see ][for an introduction to XF and FX correlators]{SynthImagIIch4}.

If the antennas of an array are located on an evenly spaced grid and are identical, a direct imaging correlator may be used \citep{Daishido:2000p3323}. The direct imaging or FFT correlator (there a several names in the literature) uses the relationship that a spatial Fourier transform of the the electric field incident on the ground ($\r$) is the observed electric field as a function of celestial direction ($\sky$, where both $\r$ and $\sky$ are two dimensional vectors)
\beq
E(\sky,[f_{k},t_{j}]) = \F(\sky,\r)\,\F([f_{k},t_{j}],t_{i})\,E(\r,t_{i}),
\eeq
where we have used the regular spacing of the antennas to map from antenna number $a$ to a location on the ground $\r$. A direct imaging correlator goes directly to a model of the sky by squaring and time averaging the electric field image
\beq
I(\sky,f_{k}) = \Big<\Big|E(\sky,[f_{k},t_{j}])\Big|^{2}\Big>_{t_{j}}.
\eeq
The advantage of the direct imaging correlator is that it can use a spatial Fast Fourier Transform ($N\log_2 N$, where $N$ is the number of inputs) instead of computing all pairwise visibilities ($N^2$), providing a large computational advantage for arrays contemplating tens of thousands of antennas. To date the largest implementation is an 8x8 array at Waseda University in Japan \citep{Daishido:2000p3323}. While computationally appealing, direct imaging correlators suffer a number of significant problems including:  poor point spread function (array beam) due to the regular grid spacing of the antennas, aliasing of sources outside the field-of-view, poor calibration as all antennas are assumed identical, and challenging deconvolution as visibilities are never formed.

The MOFF correlator is able to use the computational efficiency of the spatial Fourier transform, while using recent advances in data processing \citep{Bhatnagar:2008p3407, MoralesSoftHol} to create a data product that is equivalent in quality and calibration to the visibilities of an XF or FX correlator.

\section{Software holography/A-transpose}
\label{SHsec}

%\begin{figure}
%\begin{center}
%%\includegraphics[width = 6in]{300MHz_antenna_pattern_bw.pdf}
%\plotone{f1.eps}
%\caption{\footnotesize{This figure shows the contours from a dirty map for a single baseline using software holography. For this example we have used an antenna consisting of a 4x4 array of dipoles with a spacing of nearly one wavelength so there are strong grating lobes, and a very widefield image to show both the primary beam (center) and grating sidelobes (surrounding). This is similar to the beam patterns seen by LOFAR and MWA antennas near the top of their frequency bands. In the image the fringe from the single visibility is clearly seen as the diagonal corrugations, but its amplitude has been enveloped by the known antenna pattern and the sidelobes are clearly evident. In traditional interferometric analysis, the corrugations would have the same amplitude across the image, as the \emph{prior} of the antenna pattern is not used. In software holography the enveloping power pattern can vary from baseline-to-baseline to accurately represent the directional sensitivity of individual antenna pairs. Not shown here is the direction-dependent shifting of the fringe peaks which can be produced by directional differences in the phase delays of the antennas. While the beam pattern covers the sky, the corresponding $uv$ kernel is very compact.} }
%\label{MBEmap}
%\end{center}
%\end{figure}

Recent papers by \citet{MoralesSoftHol} and \citet{Bhatnagar:2008p3407} have introduced the software holography/A-transpose approach for analyzing interferometric radio data. These extend the optimal map making formalism widely used for the analysis of CMB data \citep{Tegmark:1997p2009,Tegmark:1997p2012} to interferometric data sets. \citet{Bhatnagar:2008p3407} have shown software holography achieves an order-of-magnitude increase in dynamic range for VLA imaging of crowded fields, and even more impressively thermal noise limited residuals of widefield polarimetric measurements (Bhatnagar, private communication).

While a full review of software holography is beyond the scope of this paper, software holography can be conceptualized as changing the post-correlation processing step of gridding visibilities to the \uv-plane from using spatial delta-functions (convolved with anti-aliasing kernels) to gridding with the holographic antenna power patterns. 

Mathematically the instrumental measurement of the visibilities---what the interferometer sees---can be written as
\begin{subequations}
\beq
\label{measEq}
%\m(\v) = \underbrace{\Sv(\v,\u)}_{4}\underbrace{\F(\u,\sky)}_{3}\underbrace{\B(\sky,\sky)}_{2}\underbrace{I(\sky)}_{1} + \underbrace{\n(\v)}_{5}.
\m(\v) = \Sv(\v,\u)\F(\u,\sky)\B(\sky,\sky)I(\sky) + \n(\v).
\eeq
or equivalently in integral notation
\beq
\label{intMeasEq}
v_i = \int \delta(\u-\u_i)\left[\int e^{-i2\pi\,\u\cdot\sky} B(\sky)I(\sky) d^2\sky \right] d^2\u + n_i.
\eeq
\end{subequations}
Reading through the steps of Equation \ref{measEq}, we start with the sky brightness $\I(\sky)$ and multiply by the antenna power response $\B(\sky,\sky)$. We can then Fourier transform directional coordinates ($\sky$) to \uv-coordinates (\u) which are then sampled at the locations of the baselines ($\Sv$) and added to the receiver noise ($\n$) to give a vector of the measured visibilities $\m(\v)$.  We can re-express Equations \ref{measEq}--\ref{intMeasEq} by Fourier transforming the antenna power response to \uv-coordinates to obtain
\bseq
\m(\v) &= &\Bv(\v,\u)\F(\u,\sky)I(\sky) + \n(\v)  \ \ \ {\rm or\ equiv.} \label{measEqU} \\
v_i &= &\int \delta(\u'-\u_i)\bigg[ \int B(\u' - \u) \times \\ 
&& \hspace{.25cm} \left[\int e^{-i2\pi\,\u\cdot\sky}I(\sky) d^2\sky \bigg]d^2\u \right] d^2\u' + n_i, \nonumber
\eseq
where in the linear algebra version we have combined the antenna power response with the delta-function sampling function ($\Sv$). Conceptually the antenna response multiplication in image space ($\sky$) has been replaced with an equivalent convolution in \uv-coordinates. The \uv-space antenna power response $\Bv(\v,\u)$ is given by convolving the $u,v$ representation of the direction-dependent antenna gains $\W(\u,\u)$,
\beq
\Bv(\v,\u) = \Sv(\v,\u)\left [ \W(\u,\u) * \W^{*}(\u,\u) \right ].
\label{Bdef}
\eeq
 The direction dependent antenna gains $\W(\u,\u)$ are just the holographic antenna maps.

Using the mathematics of CMB optimal map making \citep{Tegmark:1997p2009,Tegmark:1997p2012} a software holography image is formed using the following procedure \citep{MoralesSoftHol}:
 \beq
 \label{RBeamEq}
 \I'(\sky) = \underbrace{\F^{T}(\sky,\u)}_{3}\underbrace{ \Bv^{T}(\u,\v)} _{2}\underbrace{ \N^{-1}(\v,\v)}_{1}\m(\v).
 \eeq
 In the first step of this analysis the visibilities $\m(\v)$ are weighted by the inverse of the system noise so that high signal-to-noise channels receive more weight. The second step then grids the visibilities to a regularly sampled \uv-plane using the holographic antenna power response as the gridding kernel (and anti-aliasing filters as needed), followed by a Fourier transform in step three to create an image. The only difference between this software holography analysis and traditional imaging is the use of the antenna power pattern as the gridding kernel in step 2. 
 
In software holography the fringe from each baseline is enveloped by the power response of the two antennas used to measure that visibility. This ensures that a fringe is not reconstructed where the antennas were not sensitive, correctly adjusts fringe peaks for direction-dependent phase delays, and guarantees optimal variance weighted reconstruction at each sky pixel. The key gridding step is computationally efficient because the antenna response function in \uv-coordinates is limited to the size of the antenna---naturally truncating the size of the gridding kernel. Other effects such as w-projection, widefield distortions, and refractive and scintillating atmospheres can be easily incorporated into the software holography analysis in Equation \ref{RBeamEq}, and the interested reader is referred to \citet{MoralesSoftHol} for details.

A number of nice features can be proved about the dirty image produced by software holography, the most important of which is that the image is \emph{lossless}---all of the sky information contained in the visibilities has been preserved in the software holography image \citep{Tegmark:1997p2009}. Deconvolution can be performed directly on the software holography image---as all of the information in the visibilities has been preserved---and the result will be equivalent in quality to a visibility based deconvolution. For an imaging correlator targeting radio cosmology observations this is a key result:  if we can produce the software holography image described in equation \ref{RBeamEq}, it is equivalent in quality to the visibilities from a traditional FX or XF correlator.

\section{The MOFF correlator algorithm}
\label{MOFF}

The visibilities at a single frequency produced by an FX or XF correlator are:
\beq
\m(\v) = \left<E'(\a)*E'^*(\a')\right>_t,
\label{visDef}
\eeq
where $E'$ is a vector of the digitized RF streams from each antenna $(\a)$, and the convolution operator ($*$) acts between all of the antenna pairs to create the cross-power products which are time averaged to produce the measured visibilities.

The digitized electric field seen by the correlator in Equation \ref{visDef} can be described by
\bseq
%E'(\a) &=& \Sv(\a,\r)\F(\r,\sky)\W(\sky,\sky)E(\sky), \\
E'(\a) &=& \W(\a,\r)\F(\r,\sky)E(\sky) + \n(\a), \ \ \  {\rm or} \label{Ea}\\
E'_a & = & \int \delta(\r' - \r_a) \bigg[ \int W_a(\r' - \r) \times \\ 
&& \hspace{.25cm} \left[ \int e^{-i2\pi\,\r\cdot\sky}E(\sky) d^2\sky \bigg] d^2\r  \right] d^2 \r' + n_a, \nonumber
\eseq
where $E$ is the true incident electric field and $E'$ is the digitized RF stream produced by an antenna.\footnote{Quantization noise from the A/D converter is included in the `thermal' noise $\n$.} Equation \ref{Ea} is directly analogous to Equation \ref{measEqU}, where again we have taken the Fourier transform of the position dependent gain $W_a(\sky)$ and expressed it as a convolution in location $\r$ (the operator $\W(\a,\r)$ incorporates both the convolution and $\delta$-function sampling operations). This can be checked by showing that the correlation of electric fields in Equation \ref{Ea} reproduces Equation \ref{measEqU}.

The goal of the MOFF correlator is to create an image that is equivalent to the software holography image in Equation \ref{RBeamEq}, including all of the gridding and imaging steps. Starting with the software holography description and the above descriptions of the measured electric field and correlation process we can obtain the basic the MOFF imaging correlation algorithm:
\bseq
 \I'(\sky) &=& \F^{T}(\sky,\u) \Bv^{T}(\u,\v) \N^{-1}(\v,\v)\m(\v)  \label{m1}\\ 
& = &\F^T(\sky,\u)\Bv^T(\u,\v)\N^{-1}(\v,\v)\Big<E'(\a)* E'^*(\a')\Big>_t  \label{m2} \\
%& = & \F^T(\sky,\u)\B^T(\u,\v)\Big<\n^{-1}(\a)E(\a)* \n^{-1}(\a')E^*(\a')\Big>_t   \label{m3}\\
& = & \F^T(\sky,\u)\Big<\W^T(\r,a)\Ne^T(\a,\a)E'(\a)*\W^{T*}(\r',a')\Ne(\a',\a')E'^*(\a')\Big>_t \label{m4} \\
& = & \Big<\F^T(\sky,\r)\W^T(\r,a)\Ne^T(\a,\a)E'(\a)\times \F^{T*}(\sky,\r')\W^{T*}(\r',a')\Ne(\a',\a')E'^*(\a')\Big>_t  \label{m5} \\
%& = & \Big<\Big|\F^T(\sky,\r)\W^T(\r,a)\n^{-1}(\a)E(\a)\Big|^2\Big>_t. \label{m6}
& = & \Big<\bigg|\underbrace{\F^T(\sky,\r)}_{3}\underbrace{\W^T(\r,a)}_{2}\underbrace{\Ne(\a,\a)}_{1}E'(\a)\bigg|^2\Big>_t. \label{m6}
\eseq
The first line \ref{m1} is just a repeat of the software holography imaging process from Equation \ref{RBeamEq}, where the visibilities are weighted by the thermal noise, gridded to the \uv-plane, and Fourier transformed to produce an image. In the second step (\ref{m2}) we then substitute in the definition of the measured visibilities in terms of the digitized antenna streams from Equation \ref{visDef}. Equation \ref{m4} in the third line then moves the thermal noise weighting and gridding $\B^T(\u,\v)\N^{-1}(\v,\v)$ inside the time average, breaking the operations into their component parts using Equation \ref{Bdef} and defining a new electric field noise weighting operator satisfying $\N^{-1} = \Ne^T\Ne$ (because $\N$ is Hermitian $\Ne$ is guaranteed to exist). For the typical case of independent antenna noise $\Ne$ is simply one over the electric field noise for each antenna, but  in principle $\Ne$ can handle non-diagonal noise covariances as well . Conceptually, in this line we weight by the inverse thermal noise ($\Ne$) and use the holographic antenna pattern $\W^T(\r,\a)$ to grid the electric field measurements onto a regularly spaced positional grid $\r$, with the convolution now operating over the positional grid to produce gridded \uv-plane measurements. In the final steps  we pull the Fourier transform into the time average (\ref{m5}) to convert the convolution to a multiplication and simplify into our final algorithm in line \ref{m6}.

\begin{sidewaysfigure*}
\plotone{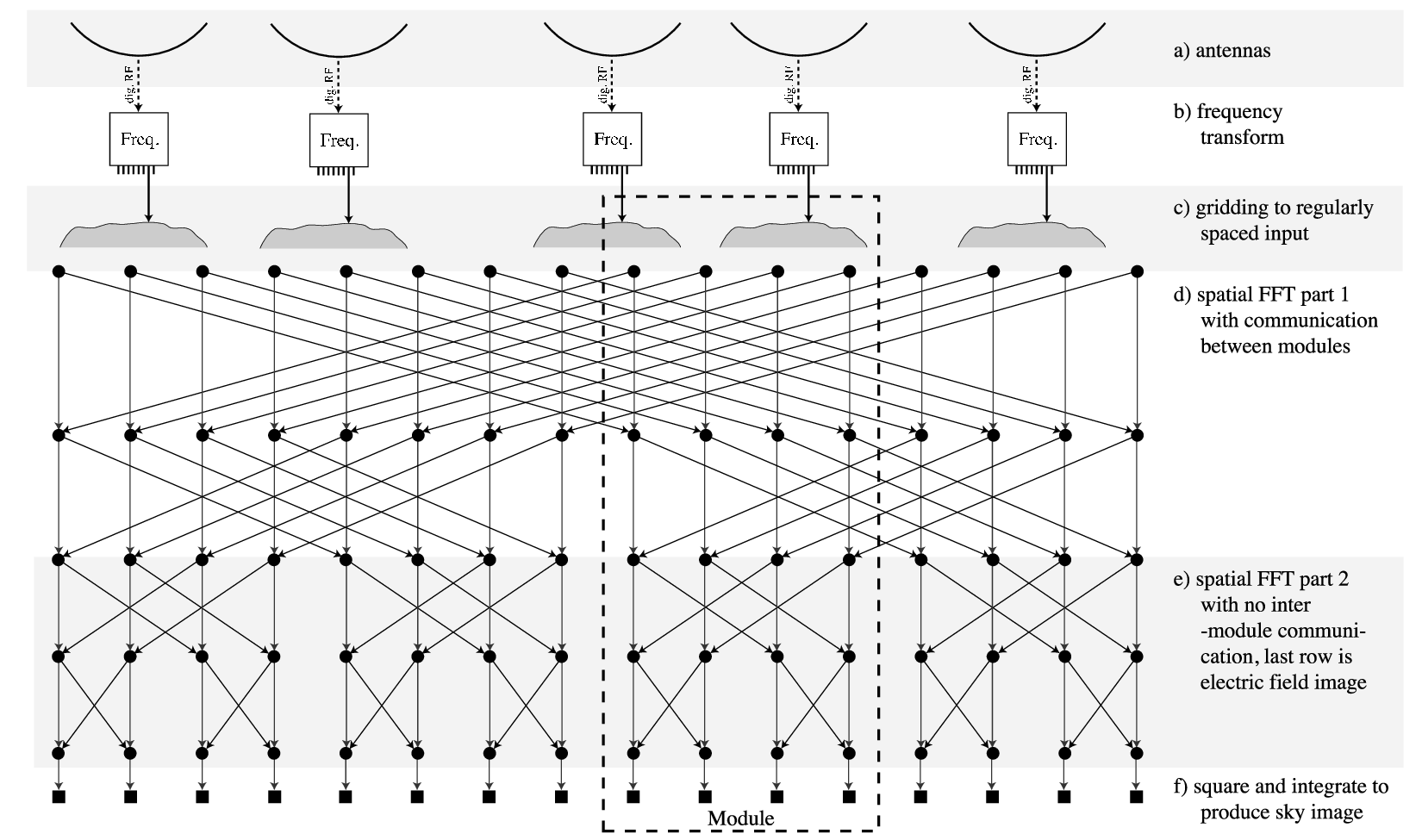}
\caption{\footnotesize{This cartoon graphically depicts a one-dimensional version the MOFF correlator structure, labeled by rows a--f (right hand side).  In row a) the electric field gathered by a 1D array of irregularly spaced antennae ($E(\a)$, Equation \ref{Ea}) is digitized and sent to the frequency portion of the correlator. The digital antenna signals are then transformed into many narrow frequency channels in row b). Through this step the MOFF correlator is identical to an FX correlator. In step c) the antenna signals are weighted by the thermal noise and gridded onto a regularly spaced grid of input nodes using the holographic antenna patterns. This step is analogous to the \uv-gridding procedure in a standard imaging pipeline. The spatial FFT algorithm is shown in rows d) and e), illustrated as a power of two Cooley-Tukey algorithm \citep{FFTbook}. At each node in the algorithm the two numbers indicated by the input arrows are multiplied by a rotation and summed. The output at the last node of row e) is a spatial image of the electric field (however the pixel locations have been scrambled by the algorithm). In the last row f), the electric field image is squared and integrated to create the interferometric image. The FFT algorithm can be separated into spatially separate `modules' as indicated by the heavy dashed box (\S \ref{modular}). Calculating the stages of the FFT within row d) requires communication between the modules, whereas the portion of the FFT within row e) is wholly contained within one module.}}
\label{moffFig}
\end{sidewaysfigure*}

Conceptually, the digitized electric field from each antenna $E'$ is weighted by the thermal noise (step 1), gridded to a regularly spaced positional grid $\r$ using the holographic antenna gain pattern (step 2, including anti-aliasing filters as needed), and spatially Fourier transformed to produce an electric field image (step 3). The electric field image is then squared and averaged to produce the snapshot image. A graphical representation of the correlation process is shown in Figure \ref{moffFig}.

This algorithm is the basis of the MOFF correlator.  Conceptually what it has done is to take the visibility calibration and gridding steps that are normally performed as part of the post-correlation imaging pipeline, and has placed them before the input to the X part of the correlator. The resulting gridded electric field can be correlated using an efficient spatial FFT, greatly reducing the required computation for compact radio arrays. By incorporating the gridding as part of the correlation process, a MOFF correlator has a number of advantages over traditional imaging correlators:
\begin{enumerate}
  \item The antennas do not need to be placed on a regular grid. The gridding operation produces the regularly spaced input needed by a spatial FFT---the antennas can be arranged in any pattern.
  \item The grid spacing can be chosen to match the FOV and science case, using the same considerations used to determine the \uv-grid spacing in standard imaging.
  \item The antenna signal is calibrated using the holographic gain pattern so antennas of any size and type can be used, including heterogenous arrays, and naturally accounts for antenna-to-antenna variations. 
  \item The output image has the equivalent information of the FX or XF correlator visibilities, allowing precision deconvolution and polarimetry. 
\end{enumerate}

There are a number of implementation subtleties that have been worked through, but are not detailed here. For example, for while the phase center should be tracked by delaying the signal of each antenna  prior to gridding (as is usually done), it is more computationally convenient to tie the east-west orientation of the input grid to the ground and do synthesis rotation in the image plane than to rotate the grid with the sky. Other imaging effects can be readily incorporated, including w-projection for non-coplanar baselines (by adjusting the gridding kernel), widefield distortions (time dependent image coordinates), image oversampling (edge padding of input FFT), and widefield refractive and scintillating atmospheric distortions \citep{MoralesSoftHol}.

\section{Modular design}
\label{modular}

A key step in any correlator is parallelizing the computation. For the MOFF correlator one could use the same approach as an FX correlator and parallelize by frequency. In this approach one X board handles all of the antennas (or spatial nodes) for a small range of frequencies. The complication is the `corner turn' between the F and X stages:  the output of the F stage boards is all the frequencies for one antenna, and the input of the X stage boards is all the antennas over a small subset of the frequencies. Much of the design of modern FX correlators is concerned with efficiently performing this corner turn to allow the necessary parallelization \citep[e.g.\ ][]{Parsons:2008p3266}.

Due to the structure of the spatial FFT, there is an alternate approach to parallelizing the computation of the MOFF correlator which may offer some distinct advantages. The spatial FFT can be separated into spatial units we are calling `modules,' as shown in Figure \ref{moffFig}.  Each module would calculate a portion of the full FFT for all of the frequencies. For the first few stages of the FFT this approach will require horizontal communications between the modules to pass intermediate results through the array (row d). The later stages of the FFT are all contained within one module (row e) and can be performed without additional module-to-module communication.

In the module approach to parallelization, inter-module communication is required at several different stages. In the gridding stage (row b), the kernel used to grid can overlap the boundary of adjacent modules. In this case a copy of the digital antenna signal needs to be provided to the neighboring module(s). The gridding kernels are usually about the size of the antennas in extent, so the fraction of antennas which straddle module boundaries tends to be small. Inter-module communication is also required in the first stages of the spatial FFT where intermediate solutions must be exchanged, as shown in Figure \ref{moffFig}. The ratio of communication to computation can be optimized by adjusting the portion of the array handled by one module. The computational requirements of a module grow in relation to the area of the array handled (number of input nodes), while the communication requirements grow as the circumference of that area.\footnote{There are also mixed-mode parallelization options, where the correlator is parallelized by frequency within one module (multiple boards) but broken down spatially across the array, with the advantage of a much smaller corner turn.}

Breaking the correlator into spatial modules could significantly reduce the cost of radio cosmology arrays. In costing out radio cosmology arrays, the cost of running cable from all the antennas to a central correlator is surprisingly large and can often drive the array design. In the MWA the antenna signals are digitized and multiplexed onto a small number of digital fibers near the antenna, and similar approaches are used for LOFAR and the Long Wavelength Array (LWA\footnote{\url{http://www.phys.unm.edu/~lwa/index.html}}). For the MOFF correlator we could include the correlator modules in the field-based digital receivers, so the correlator can distributed throughout the array. This would reduce the cabling to only neighbor-to-neighbor connections between receiver/correlator modules, eliminating the long antenna to central correlator runs.

In addition to cost savings, this could help the phased build-out of an array. As each section of the array is completed and its correlator module is connected to its neighbors, it immediately enhances the scientific reach of the observatory. Thus an array could be built in stages over a number of years, gradually increasing in capability. This growth of capability is much more difficult for an FX correlator where the corner turn is optimized for a particular number of antennas.

The modularity of the MOFF correlator does not come without a cost, as there is significant latency in inter-module communications. These delays can be hidden by pipelining the computation so other correlation work is performed while the intermediate FFT products are en route. This pipelining and communication architecture is a major area of continuing development.

\section{Feedback Calibration}
\label{calibration}

While software holography describes how to precisely calibrate the data (\S\ref{SHsec}), it is difficult to determine what this calibration should be with imaging correlators. In traditional data processing self-cal or related algorithms use  visibilities to determine the calibration during post processing, but imaging correlators never calculate the visibilities self-cal needs. The MOFF correlator does not form visibilities either, however it is possible to determine the calibration by correlating pixels of the electric field image with the antenna signal---feeding the correlator output back to the input.

The electric field image (last line of  row $e$ in Figure \ref{moffFig}) in the direction of a calibrator source is the sum of the electric field from all of the antennas towards that source. So the square of the electric field towards the calibrator is equivalent to the sum of all the calibrated visibilities summed towards the calibrator, as expected
%\beq
%\left<|E(\theta_{\rm cal})|^2\right>_t = \left<\left(\sum_{i} g^{*}_{M}\, g_{T} E_{i}\right) \left(\sum_{j} g^{*}_{M}\, g_{T} E_{j})\right)^*\right>_t = 
%\sum_{ij} (g_M^*g_T)_i(g_M^*g_T)_j\,v_{ij}
%\label{FBeq1}
%\eeq
\beq
\left<|E(\theta_{\rm cal})|^2\right>_t = \left<\left(\sum_{i} g_i E_{i}\right) \left(\sum_{j} g_j^* E_{j}^*\right)\right>_t = 
\sum_{ij} g_ig^*_j\,v_{ij},
\label{FBeq1}
\eeq
where we are implicitly summing in the direction of the calibrator.
%\beq
%E(\theta_{\rm cal}) = \sum_{a} g^{*}_{M}(\theta_{\rm cal})\, g_{T}(\theta_{\rm cal}) E_{a}
%\eeq
If we instead correlate the pixel of the electric field image towards the calibrator  with the calibrated signal of one antenna---correlating the output of the correlator with the input---we obtain the sum of the just the visibilities involving that antenna
%\beq
%\big<E(\theta_{\rm cal})\,g_M^*g_TE^*_{i}\big>_{t} = \left<\left(\sum_{j} g^{*}_{M}\, g_{T} E_{j}\right)g_M^*g_TE_i^*\right>_t = \sum_{j} (g_M^*g_T)_i(g_M^*g_T)_j\,v_{ij}
%\label{FBeq2}
%\eeq
\beq
\big<E(\theta_{\rm cal})\,g_i^*E^*_{i}\big>_{t} = \left<\left(\sum_{j} g_j E_{j}\right)(g_i^*E_i^*)\right>_t = \sum_{j} g_ig_j^*\,v_{ij}.
\label{FBeq2}
\eeq
We can also write down the sum of all visibilities not including the antenna in question
\beq
\sum_{jk \neq i} g_j g^*_k\,v_{jk} = \left<|E(\theta_{\rm cal})|^2\right>_t - \big<E(\theta_{\rm cal})\,g_i^*E^*_{i}\big>_{t}.
\label{FBeq3}
\eeq

We have still not measured all of the visibilities independently, but these sums are sufficient to determine the self-cal solutions towards the calibrator(s) following \cite{2008ISTSP...2..707M}. One can compare the observed brightness and phase of the calibrator with the expected value to update the calibration solution. This can be done for a single dominant calibrator, a multi-source calibration solution, or for calibrators distributed across the field-of-view to determine the holographic antenna patterns for each antenna. As an example, a good estimate of the antenna gain error in the single calibrator case is
\beq
\Delta \hat{g}_{i} \approx g_i \left(1 -  \frac{ 2N \big<E(\theta_{\rm cal})\,g_i^*E^*_{i}\big>_t}{\left<|E(\theta_{\rm cal})|^2\right>_t}\right).
\eeq
This is effectively the normalized difference between the sum of all the other antennas towards the source (Eq. \ref{FBeq3}) and the sum of the visibilities associated with the antenna (Eq. \ref{FBeq2}). The relative calibration of one antenna to the rest of the array may be converted into an absolute calibration with addition information, either in the form of an absolutely calibrated reference antenna (e.g.\ from antenna range measurements), or via absolute flux values for the astrophysical calibrators.

Using these visibility sums has been shown to work robustly for the 32 tile MWA prototype and to quickly converge on holographic antenna patterns in simulation for the 512 tile MWA \citep{2008ISTSP...2..707M}. This ability to generate calibration solutions by feeding back the output of the correlator to the input is a major advantage of the MOFF design.

%To illustrate how to solve the self-cal problem, imagine that all of the antennas except the one in question are correctly calibrated. Then we can use Equations \ref{FBeq1} \& \ref{FBeq2} to estimate the error in the gain:
%\beq
%\Delta \hat{g}_{i} \approx  \frac{ \frac{1}{N^2}\left<|E(\theta_{\rm cal})|^2\right>_t -  \frac{2}{N}\big<E(\theta_{\rm cal})\,g_i^*E^*_{i}\big>_t}{ \frac{1}{g_i N^2}\left<|E(\theta_{\rm cal})|^2\right>_t}
%\label{FBsoln}
%\eeq
%or 

%form the difference between the sum of visibilities involving just the one antenna (Eq. \ref{FBeq2}) and the normalized sum of all the antennas.
%\beq
%(g_M - g_T)_i = 
%\eeq
%\beq
%\frac{1}{N}\big<E(\theta_{\rm cal})\,g_M^*g_TE^*_{i}\big>_{t} - \frac{1}{N^2}\left<|E(\theta_{\rm cal})|^2\right>_t  = 
%\eeq

%\section{Computational comparison}
%\label{computation}

\section{Discussion}
\label{MOFFreview}

The MOFF correlator can be orders of magnitude more efficient than traditional FX correlators for very compact ultra-large N arrays. Both the FX and MOFF correlators start with a Fourier transform in the frequency domain, with the spatial correlation then proceeding within the narrow frequency channels. This first temporal Fourier transform is identical for both FX and MOFF correlators, and is typically much smaller than the spatial part of the correlation (we will follow tradition and ignore the computation in the spectral transform). For an FX correlator the number of complex multiplications and additions in the spatial part of the calculation (X) is equal to the bandwidth times the number of antennas $N_a$ squared:
\beq
{\rm FX} \propto B_v N_a^2.
\eeq
(We have indicated bandwidth with $B_v$ to avoid confusion with the antenna power response pattern referred to as $B$ earlier.)
 For the MOFF correlator the two dimensional spatial FFT scales with the number of grid locations $N_g$:
 \beq
 {\rm MOFF} \propto c \, B_v \left[ N_g \log_2 \sqrt{N_g} \right].
 \eeq
(The square root comes from assuming a square grid with the length of one side equal to $\sqrt{N_g}$, and the constant $c$ ranges between $1/2$ and 1 depending on the implementation of twiddle FFT algorithms \citep{FFTbook}.) The MOFF correlator depends only on the size of the array (and \uv\ sampling density)---adding more antennas does not change the correlation load.

The parameter space of array size and antenna density over which the MOFF correlator is more efficient than an FX correlator can be described following an argument by Aaron Parsons (personal communication). If the array filling factor $\Fill$ is defined so the number of antennas is $\Fill N_g$, the MOFF correlator will be more efficient if
\beq
 N_g \log_2 \sqrt{N_g} \lesssim \left ( \Fill N_g \right ) ^2.
\eeq
Ignoring the logarithm as a relatively small term, we can rearrange to obtain the criteria that the filling factor 
\beq
\Fill \gtrsim \frac{1}{\sqrt{N_g}}
\label{Eff1}
\eeq
for the MOFF correlator to be more efficient. As the grid spacing is often chosen to cover the primary response of the antennas, the number of grid points is typically the square of the ratio of the array diameter $(D_A)$ to the antenna diameter $(d_a)$. Substituting we obtain the criteria
\beq
\Fill \gtrsim \frac{d_a}{D_A}.
\label{Eff2}
\eeq
Equations \ref{Eff1} and \ref{Eff2} both show that as the array becomes larger in size, a smaller filling factor is needed for the MOFF correlator to be a viable alternative. An FX correlator will be more efficient for sparse arrays, because it does not calculate any correlations where there are no antennas. However, as the array filling factor becomes large the computational efficiency of the FFT starts to give the MOFF correlator a significant advantage. This means that the FX correlator is much more efficient for sparse arrays such as the VLA, but is less efficient for future cosmology focused arrays such as the HERA II, LARC, FFT Telescope, and possibly the SKA (Square Kilometer Array\footnote{\url{http://www.skatelescope.org/}}). 

For the MWA with 500 antennas in a 1.5 km array, the FX correlator under construction will perform 15.5 Tcmacs (trillions of complex multiply and additions per second, 31 MHz x 125000 baselines x 4 polarizations). An equivalent MOFF correlator would require 4-5 times as many calculations.  However, for the proposed MWA upgrade to HERA II with ten times the number of antennas in the same area, the MOFF correlator would require approximately 20 times less computation than an FX correlator. For the Lunar Array for Radio Cosmology (LARC), the MOFF correlator would be more than two orders-of-magnitude more efficient than an FX design.

%The comparison to a direct imaging correlator is trickier, because the direct imaging correlator requires a filled aperture with the antennas on a regularly spaced rectangular grid. For a filled regularly spaced array the computational efficiency is nearly identical between the direct imaging correlator and the MOFF correlator, though the image produced by the MOFF correlator is markedly superior. With the MOFF correlator the spacing of the grid is independent from the antenna spacing, and can be adjusted to optimize the FOV of the resulting image. In many cases this can reduce the number of input nodes, resulting in a more efficient calculation. However, the real advantage of the MOFF correlator is the ability to place the antennas in any arrangement and the ability to accurately calibrate data from large antenna elements that span many wavelengths.
%
%
%
%
%The MOFF correlator concept has a number of advantages over traditional correlator designs for envisioned radio cosmology observatories. The advantages we have detailed include:
%\begin{itemize}
%  \item Calibrated output image, equivalent in quality to FX visibilities (\S\ref{SHsec}).
%  \item More efficient than FX or XF correlators for compact arrays with very large numbers of antennas, sometimes by orders-of-magnitude (\S\ref{computation}).
%  \item No constraints on antenna placement or type (\S\ref{MOFF}).
%  \item Modular installation and reduced cabling costs (\S\ref{modular}).
%\end{itemize}
A couple of features we have not discussed but which follow directly from the MOFF design are:
\begin{itemize}
  \item Electric field image can be tapped for pulsar studies and transient detection. Each pixel is effectively a calibrated tied-array beam.
  \item Hybrid modes are possible, where a MOFF correlator is used for all of the antennas in a compact station and the electric field image from selected pixels is sent to an FX correlator for high angular resolution interferometry between the stations. 
\end{itemize}

As with any design there are some disadvantages as well. The primary disadvantage of the MOFF correlator is that the calibration must be applied as part of the correlation process---one cannot re-calibrate the data during post-processing. For very large cosmology arrays the overwhelming number of antenna pairs may make saving raw visibility data impractical anyway. The MWA realtime system is designed to allow the visibilities to be discarded to save storage, and future cosmology arrays will probably do the same. However, this may limit the usefulness of the MOFF concept in certain applications. 

%A second disadvantage is that the holographic antenna gain patterns needed for calibration (\S\ref{SHsec}) cannot be obtained directly from the image produced by the MOFF correlator.  Effectively the antenna beam patterns have been co-added during the correlation process (most easily seen in Equation \ref{m1}), and one needs the individual visibilities to measure the holographic antenna gain response. As part of real world implementations of the  MOFF correlator, digitized signals from selected antennas will need to be pulled out of the MOFF correlator (Figure \ref{moffFig} row b) so holographic gain patterns can be formed offline.

\section*{Acknowledgments}

Morales would like to thank Aaron Parsons, Peter Ford, Bryna Hazelton, Adam Beardsley, and Dan Werthimer for their helpful comments and suggestions, and Roger Cappallo for teaching me the art of correlator building. This work has been partially supported via NSF-AST grant \#0847753.

\end{document}